\documentclass[twoside,english,preprint,amsmath,nofootinbib]{revtex4}
\usepackage[T1]{fontenc}
\usepackage[latin9]{inputenc}
\usepackage{amsmath}
\usepackage{amssymb}
\usepackage{esint}

\makeatletter
\@ifundefined{textcolor}{}
{%
 \definecolor{BLACK}{gray}{0}
 \definecolor{WHITE}{gray}{1}
 \definecolor{RED}{rgb}{1,0,0}
 \definecolor{GREEN}{rgb}{0,1,0}
 \definecolor{BLUE}{rgb}{0,0,1}
 \definecolor{CYAN}{cmyk}{1,0,0,0}
 \definecolor{MAGENTA}{cmyk}{0,1,0,0}
 \definecolor{YELLOW}{cmyk}{0,0,1,0}
 }

\@ifundefined{definecolor}{\@ifundefined{definecolor}{\@ifundefined{definecolor}{\@ifundefined{definecolor}{\@ifundefined{definecolor}{\@ifundefined{definecolor}{\@ifundefined{definecolor}{\@ifundefined{definecolor}{\@ifundefined{definecolor}{\@ifundefined{definecolor}{\@ifundefined{definecolor}{\@ifundefined{definecolor}{\@ifundefined{definecolor}{\@ifundefined{definecolor}
 {\usepackage{color}}{}
}{}}{}}{}}{}}{}}{}}{}}{}}{}}{}}{}}{}}{}\usepackage[dvipdfm,bookmarks=true]{hyperref}

\makeatother

\usepackage{babel}

\makeatother

\usepackage{babel}

\makeatother

\usepackage{babel}

\makeatother

\usepackage{babel}

\makeatother

\usepackage{babel}

\makeatother

\usepackage{babel}

\makeatother

\usepackage{babel}

\makeatother

\usepackage{babel}

\makeatother

\usepackage{babel}

\makeatother

\usepackage{babel}

\makeatother

\usepackage{babel}

\makeatother

\usepackage{babel}

\makeatother

\usepackage{babel}

\makeatother

\usepackage{babel}

\makeatother

\usepackage{babel}

\makeatother

\usepackage{babel}

\makeatother

\usepackage{babel}
\begin{document}

\title{{\LARGE Identities in Nonlinear Realizations of Supersymmetry}}

\date{\today}

\author{Haishan Liu}

\author{Hui Luo}

\author{Mingxing Luo}

\author{Liucheng Wang}

\email{liuchengwang@gmail.com}

\selectlanguage{english}%

\thanks{(Corresponding Author)}

\affiliation{Zhejiang Institute of Modern Physics, Department of Physics, Zhejiang
University, Hangzhou, Zhejiang 310027, P.R.China}
\begin{abstract}
In this paper, we emphasize that a UV SUSY-breaking theory can be realized either linearly or nonlinearly.
Both realizations form the dual descriptions of the UV SUSY-breaking theory.
Guided by this observation, we find subtle identities involving the
Goldstino field and matter fields in the standard nonlinear realization 
from trivial ones in the linear realization.
Rather complicated integrands in the standard nonlinear realization are
identified as total-divergences. 
Especially, identities only involving the Goldstino field reveal 
the self-consistency of the Grassmann algebra. 
As an application of these identities, we prove that the nonlinear Kahler potential without or with gauge interactions is unique, if the corresponding linear one is fixed.
Our identities pick out the total-divergence terms and guarantee this uniqueness.
\end{abstract}

\maketitle

As a possible extension of the Standard Model, supersymmetry (SUSY) has attracted
much attention over several decades.
SUSY has the important feature that it provides a reasonable framework to circumvent the
hierarchy problem.
One may start the discussion of SUSY theories from the SUSY algebra, which is
the only viable graded Lie algebra consistent with other requirements for a non-trivial relativistic
quantum field theory. For any field $A$, the N=1 SUSY algebra requires 
(for review, see \cite{WessBagger},\cite{West},\cite{Martin:1997ns})
\begin{equation}
\left(\delta_{\eta}\delta_{\xi}-\delta_{\xi}\delta_{\eta}\right)A=-2i\left(\eta\sigma^{\mu}\bar{\xi}-\xi\sigma^{\mu}\bar{\eta}\right)\partial_{\mu}A,\label{eq:closure relation}
\end{equation}
where the infinitesimal SUSY transformation $\delta_{\xi}$ is realized
by the operator $\xi Q+\bar{\xi}\bar{Q}$ as
\begin{equation}
\delta_{\xi}A=\left(\xi Q+\bar{\xi}\bar{Q}\right)A.\label{eq:infinitesimal}
\end{equation} 
Much has been discussed for SUSY models and phenomenological implications.
Hopefully, one is to see some SUSY signals in the running LHC experiment.

In order to be consistent with existing experimental
measurements and to be of phenomenological relevance,
SUSY must be broken and broken spontaneously at the TeV scale.
According to the general theory of spontaneously symmetry breaking, this should
lead to a massless neutral Nambu-Goldstone fermion, the Goldstino field. 
To deal with the low energy SUSY physics,
it is expedient to study fields in the framework of nonlinear realization of SUSY. 
In the nonlinear realization, vertices with Goldstino fields always carry at least one space-time derivative, as one would have expected. 
So it is difficult to directly pick out the total-divergence terms in the nonlinear actions.
In the Appendix 2 of \cite{Ivanov:1982bpa}, 
a rather complicated integrand related to Goldstino field was proven to be total-divergence via tedious Fierz rearrangements. 
This total-divergence term simplifies the non-minimal contribution of the Goldstino action. 
This feature brings us to study how to identify more total-divergence terms in the nonlinear SUSY actions.

In this paper, we observe more identities in the nonlinear SUSY actions. 
We emphasize that a UV SUSY-breaking theory could be realized either linearly or nonlinearly. 
Both realizations form dual descriptions of the same UV SUSY-breaking theory.
Keeping the dual description in mind, 
the total-divergence terms in a linear SUSY theory can be transformed into their nonlinear versions. 
We thus obtain intriguing identities about the Goldstino field and matter fields 
in the standard nonlinear realization. 
Rather complicated integrands are identified as total-divergences. 
One of these identities is exactly the one in the Appendix 2 of \cite{Ivanov:1982bpa} and other identities are first obtained. Moreover, we show that the identity in the Appendix 2 of \cite{Ivanov:1982bpa} reveals the self-consistency of the Grassmann algebra.
As an application of our identities, 
we prove presently the equivalence between some different kinds of nonlinear Kahler potentials.
The nonlinear Kahler potential without or with gauge interactions is unique, 
if the corresponding linear one is fixed.
Our identities pick out the total-divergence terms and guarantee this uniqueness.

Primarily, we will discuss the standard nonlinear realization as first introduced in \cite{Volkov:1973ix}.
SUSY transformations of the Goldstino filed $\lambda$ and the matter field $\zeta$ are as follows \cite{WessBagger}
\begin{equation}
\begin{cases}
\delta_{\xi}\lambda_{\alpha} & =  \frac{\xi_{\alpha}}{\kappa}-i\kappa\left(\lambda\sigma^{\mu}\bar{\xi}-\xi\sigma^{\mu}\bar{\lambda}\right)\partial_{\mu}\lambda_{\alpha}, \label{eq:nl-transformation} \\ 
\delta_{\xi}\zeta & =  -i\kappa\left(\lambda\sigma^{\mu}\bar{\xi}-\xi\sigma^{\mu}\bar{\lambda}\right)\partial_{\mu}\zeta.  
\end{cases}
\end{equation}
It is easy to check that both equations in Eq (\ref{eq:nl-transformation}) satisfy the closure relation Eq (\ref{eq:closure relation}). 
This nonlinear realization could be constructed from the linear one
with the help of the standard realization of the nonlinear SUSY 
\cite{Ivanov:1982bpa},\cite{Ivanov:1977my},\cite{Ivanov:1978mx}.
It is based upon the following observation: 
with the help of the nonlinear Goldstino field $\lambda$, 
a linear superfield $\hat{\Omega}(x,\theta,\bar{\theta})$ can be converted into
a set of nonlinear matter fields via\footnote{In this paper, superfields and their components in the linear SUSY
are hatted while their counterparts in the nonlinear SUSY are not. We use the conventions of \cite{WessBagger}. 
All symbols can be found in \cite{Luo:2010zp}, if not explicitly defined in this paper.}
\begin{equation}
\Omega(x,\theta,\bar{\theta})=\exp\left[-\kappa\lambda(x)Q-\kappa\bar{\lambda}(x)\bar{Q}\right]\times\hat{\Omega}(x,\theta,\bar{\theta}),\label{eq:promotion}
\end{equation}
where $Q_{\alpha}$ and $\bar{Q}_{\dot{\alpha}}$ are generators of SUSY algebra realized in superspace
\begin{equation}
\left\{ \begin{array}{l}
Q_{\alpha}=\partial_{\alpha}+i\left(\sigma^{\mu}\bar{\theta}\right)_{\alpha}\partial_{\mu},\\
\bar{Q}_{\dot{\alpha}}=-\partial_{\dot{\alpha}}-i\left(\theta\sigma^{\mu}\right)_{\dot{\alpha}}\partial_{\mu}.
\end{array}\right.
\end{equation}
Since $\lambda$ transforms according to the first equation in Eq (\ref{eq:nl-transformation})
and $\hat{\Omega}(x,\theta,\bar{\theta})$ transforms according to
Eq (\ref{eq:infinitesimal}), it is easy to prove that $\Omega(x,\theta,\bar{\theta})$
transforms according to the second equation in Eq (\ref{eq:nl-transformation}).
Following these redefinitions of fields, 
any UV actions in linear SUSY-breaking theories could be re-expressed in the language of nonlinear fields \cite{Ivanov:1982bpa},\cite{Luo:2010zp}. 
This enables us to construct dual descriptions of the same UV SUSY-breaking theory, which could be described either linearly or nonlinearly.
Both realizations have the same degrees of freedom. 
The difference are in definitions of fields, which respect the SUSY algebra in their own ways.
Both realizations can be connected by the so-called standard realization of the nonlinear SUSY 
\cite{Ivanov:1982bpa},\cite{Ivanov:1977my},\cite{Ivanov:1978mx}.
To show this duality, we take the UV canonical Kahler potential as an example. 
For a chiral superfield $\hat{\Phi}$ in a linear theory, the canonical Kahler potential is 
${\cal S}_{{\rm can}}=\int d^{4}xd^{4}\theta\hat{\Phi}^{\dagger}\hat{\Phi}$. 
After fields redefinitions\footnote{In order to get SUSY-breaking, 
the F-term of the linear chiral field $\hat{\Phi}$ should develop a non-zero VEV $\left \langle F \right \rangle$.
The nonlinear Goldstino field can then be constructed from fields of the linear theory
by setting the fermionic component vanishing \cite{Luo:2009pz}.}, one has ${\cal S}_{{\rm can}}^{{\rm NL}}=\int d^{4}xd^{4}\theta\det{\emph{T}}\det{\emph{M}}\;\Phi^{\dagger}\Phi$ in its dual nonlinear description \cite{Ivanov:1982bpa},\cite{Luo:2010zp}. 
Both ${\cal S}_{{\rm can}}$ and ${\cal S}_{{\rm can}}^{{\rm NL}}$ have the same degrees of freedom and describe the same physical process.

Interestingly, the canonical Kahler potential in linear theory may be written
as ${\cal S}_{{\rm can}}^{'}=\int d^{4}xd^{4}\theta\frac{1}{2}\left(e^{-2i\theta\sigma^{\mu}\bar{\theta}\partial_{\mu}}\hat{\varphi}^{\dagger}\hat{\varphi}+\hat{\varphi}^{\dagger}e^{2i\theta\sigma^{\mu}\bar{\theta}\partial_{\mu}}\hat{\varphi}\right)$
by dropping possible total-divergence terms. 
Rewriting ${\cal S}_{{\rm can}}^{'}$ in terms of nonlinear chiral superfields \cite{Ivanov:1982bpa},\cite{Luo:2010zp}, 
we obtain ${\cal S}_{{\rm can}}^{'{\rm NL}}=\int d^{4}xd^{4}\theta\frac{1}{2}\det{\emph{T}}\left(\det{\emph{M}_{+}}e^{-2i\theta\sigma^{\mu}\bar{\theta}\triangle_{\mu}^{-}}\varphi^{\dagger}\varphi+\det{\emph{M}_{-}}\varphi^{\dagger}e^{2i\theta\sigma^{\mu}\bar{\theta}\triangle_{\mu}^{+}}\varphi\right)$. 
No doubt both ${\cal S}_{{\rm can}}$ and ${\cal S}_{{\rm can}}^{'}$ describe the same physical process.
As dual descriptions, the nonlinear versions ${\cal S}_{{\rm can}}^{{\rm NL}}$ and ${\cal S}_{{\rm can}}^{'{\rm NL}}$ should 
describe the same physics too.
However, the integrands of ${\cal S}_{{\rm can}}^{{\rm NL}}$ and ${\cal S}_{{\rm can}}^{'{\rm NL}}$ have quite different terms after integrating out the $\theta$'s.
In this paper, we will show that those different terms can be identified as total-divergences. 
As an application of our identities, we will show the equivalence 
between ${\cal S}_{{\rm can}}^{{\rm NL}}$ and ${\cal S}_{{\rm can}}^{'{\rm NL}}$ in the end of this paper.

We start our discussion from Grassmann algebra in any linear theory, which is the simplest case in our observations.
Grassmann algebra requires the identity $\int d^{4}xd^{2}\theta\times1=0$.
In its dual description, this can be re-expressed in terms of the nonlinear fields as
\begin{equation}
\int d^{4}xd^{2}\theta\det{T}\,\det{\emph{M}_{+}}\times1=0.
\end{equation}
Here $\det{T}\,\det{\emph{M}_{+}}$ is the Jacobian determinant of the transformation Eq (\ref{eq:promotion}),
with $\emph{{T}}\,_{\mu}^{\nu}=\delta_{\mu}^{\nu}-i\kappa^{2}\partial_{\mu}\lambda\sigma^{\nu}\bar{\lambda}+i\kappa^{2}\lambda\sigma^{\nu}\partial_{\mu}\bar{\lambda}$ and $\emph{{\emph{M}}}\,_{+\mu}^{\nu}=\delta_{\mu}^{\nu}-2i\kappa\theta\sigma^{\nu}\bar{\lambda}_{\mu}$  \cite{Ivanov:1982bpa}.
Explicitly,
\begin{eqnarray}
\det{T} & = & 1-i\kappa^{2}\left(\partial_{\mu}\lambda\sigma^{\mu}\bar{\lambda}-\lambda\sigma^{\mu}\partial_{\mu}\bar{\lambda}\right)\label{eq:det T}\\
 &  & -\kappa^{4}\left(i\epsilon^{\mu\nu\rho\gamma}\lambda\sigma_{\rho}\bar{\lambda}\partial_{\mu}\lambda\sigma_{\gamma}\partial_{\nu}\bar{\lambda}+\bar{\lambda}^{2}\partial_{\mu}\lambda\sigma^{\mu\nu}\partial_{\nu}\lambda+\lambda^{2}\partial_{\mu}\bar{\lambda}\bar{\sigma}^{\mu\nu}\partial_{\nu}\bar{\lambda}\right)\nonumber \\
 &  & -i\kappa^{6}\lambda^{2}\bar{\lambda}\left(\bar{\sigma}^{\rho}\partial_{\rho}\lambda\partial_{\mu}\bar{\lambda}\bar{\sigma}^{\mu\nu}\partial_{\nu}\bar{\lambda}+2\bar{\sigma}^{\nu}\partial_{\mu}\lambda\partial_{\nu}\bar{\lambda}\bar{\sigma}^{\rho\mu}\partial_{\rho}\bar{\lambda}\right)\nonumber \\
 &  & -i\kappa^{6}\bar{\lambda}^{2}\lambda\left(\sigma^{\rho}\partial_{\rho}\bar{\lambda}\partial_{\mu}\lambda\sigma^{\mu\nu}\partial_{\nu}\lambda+2\sigma^{\nu}\partial_{\mu}\bar{\lambda}\partial_{\nu}\lambda\sigma^{\rho\mu}\partial_{\rho}\lambda\right),\nonumber \\
\det{\emph{M}_{+}} & = & 1-2i\kappa\theta\sigma^{\mu}\bar{\lambda}_{\mu}+4\kappa^{2}\theta^{2}\bar{\lambda}_{\mu}\bar{\sigma}^{\nu\mu}\bar{\lambda}_{\nu},\label{eq:det M+}
\end{eqnarray}
 with $\lambda_{\mu}=\nabla_{\mu}\lambda$, $\bar{\lambda}_{\mu}=\nabla_{\mu}\bar{\lambda}$
and $\nabla_{\mu}=({T}^{-1}){}_{\mu}^{\nu}\partial_{\nu}$.
Naively, one may expect $\kappa^{8}\lambda^{2}\bar{\lambda}^{2}$ terms
in $\det{T}$ by expanding $\emph{{T}}\,_{\mu}^{\nu}$ directly.
However, this term vanishes actually \cite{Kuzenko:2005wh},\cite{Zheltukhin:2010rj},\cite{Liu:2010sk}.
Integrating out the $\theta$'s, one has
\begin{equation}
\int d^{4}x\,\det{T}\,\bar{\lambda}_{\mu}\bar{\sigma}^{\nu\mu}\bar{\lambda}_{\nu}=0.\label{eq:identity G1}
\end{equation}
Starting with $\int d^{4}xd^{2}\bar{\theta}\times1=0$, one obtains another  identity
\begin{equation}
\int d^{4}x\,\det{T}\,\lambda_{\mu}\sigma^{\nu\mu}\lambda_{\nu}=0,\label{eq:identity G2}
\end{equation}
which is the conjugate of Eq (\ref{eq:identity G1}).

Eq (\ref{eq:identity G2}) tells us that the term $\det{T}\,\lambda_{\mu}\sigma^{\nu\mu}\lambda_{\nu}$
should be a total divergence term. So its integration, which masquerades
as a non-zero result, should vanish. In fact, this conclusion was first
shown in the Appendix 2 of \cite{Ivanov:1982bpa}, where $\det{T}\,\lambda_{\mu}\sigma^{\nu\mu}\lambda_{\nu}$
was proven to be a total divergence term via tedious Fierz rearrangements.
We comment that Eq (\ref{eq:identity G2}) reveals the self-consistency
of Grassmann algebra.
One starting point of the SUSY nonlinear realization is to identify a Grassmannian coordinate $\xi$ with
the Goldstino field $\lambda$, namely $\xi\rightarrow\kappa\lambda$
\cite{WessBagger},\cite{Volkov:1973ix}. Eq (\ref{eq:promotion}) is essentially
a special shift $\theta'=\theta-\kappa\lambda(x)$ in superspace.
The shift-invariance of Grassmann algebra demands both $\int d^{4}xd^{2}\theta\times1=0$
and $\int d^{4}x'd^{2}\theta'\times1=0$, which leads to Eq(\ref{eq:identity G1}).
This integral gives no contribution to any physical process.
Its integrand would be discarded when a Lagrangian of the Goldstino field is constructed \cite{Ivanov:1982bpa}.

Grassmann algebra also requires
$\int d^{4}xd^{4}\theta\times1=0$ in a linear theory.
This can also be re-expressed in the nonlinear realization as
\begin{equation}
\int d^{4}xd^{4}\theta\det{T}\,\det{\emph{M}}\times1=0,\label{eq:identity G3}
\end{equation}
 where $\emph{{\emph{M}}}\,_{\mu}^{\nu}=\delta_{\mu}^{\nu}-i\kappa\theta\sigma^{\nu}\bar{\lambda}_{\mu}+i\kappa\lambda_{\mu}\sigma^{\nu}\bar{\theta}$ and
\begin{eqnarray}
\det{\emph{M}} & = & 1+i\kappa\left(\lambda_{\mu}\sigma^{\mu}\bar{\theta}-\theta\sigma^{\mu}\bar{\lambda}_{\mu}\right)\label{eq:det M}\\
 &  & -\kappa^{2}\left(i\epsilon^{\mu\nu\rho\gamma}\theta\sigma_{\rho}\bar{\theta}\lambda_{\mu}\sigma_{\gamma}\bar{\lambda}_{\nu}+\bar{\theta}^{2}\lambda_{\mu}\sigma^{\mu\nu}\lambda_{\nu}+\theta^{2}\bar{\lambda}_{\mu}\bar{\sigma}^{\mu\nu}\bar{\lambda}_{\nu}\right)\nonumber \\
 &  & +i\kappa^{3}\theta^{2}\bar{\theta}\left(\bar{\sigma}^{\rho}\lambda_{\rho}\bar{\lambda}_{\mu}\bar{\sigma}^{\mu\nu}\bar{\lambda}_{\nu}+2\bar{\sigma}^{\nu}\lambda_{\mu}\bar{\lambda}_{\nu}\bar{\sigma}^{\rho\mu}\bar{\lambda}_{\rho}\right)\nonumber \\
 &  & +i\kappa^{3}\bar{\theta}^{2}\theta\left(\sigma^{\rho}\bar{\lambda}_{\rho}\lambda_{\mu}\sigma^{\mu\nu}\lambda_{\nu}+2\sigma^{\nu}\bar{\lambda}_{\mu}\lambda_{\nu}\sigma^{\rho\mu}\lambda_{\rho}\right).\nonumber
\end{eqnarray}
Noticing that there is no $\theta^{2}\bar{\theta}^{2}$  term in $\det{\emph{M}}$,
just as there is no  $\kappa^{8}\lambda^{2}\bar{\lambda}^{2}$ term in $\det{T}$.
So Eq (\ref{eq:identity G3}) leads to a trivial
but self-consistent result.

From a linear chiral superfield
$\hat{\Phi}(x,\theta,\bar{\theta})=\exp(i\theta\sigma^{\mu}\bar{\theta}\partial_{\mu})\hat{\varphi}(x,\theta)$
with $\hat{\varphi}(x,\theta)=\hat{\phi}+\sqrt{2}\theta\hat{\psi}+\theta^{2}\hat{F}$,
a nonlinear $\Phi$ can be obtained via Eq (\ref{eq:promotion})
as $\Phi(x,\theta,\bar{\theta})=\exp(i\theta\sigma^{\mu}\bar{\theta}\triangle_{\mu}^{+})\varphi(x,\theta)$,
with $\varphi(x,\theta)=\phi+\sqrt{2}\theta\psi+\theta^{2}F$ and
$\triangle_{\mu}^{+}={(\emph{M}_{+}^{-1})_{\mu}}^{\nu}(\nabla_{\nu}+\kappa\lambda_{\nu}^{\alpha}\partial_{\alpha})$ \cite{Ivanov:1982bpa},\cite{Luo:2010zp}.
Here $\phi$, $\psi$ and $F$ are composites of $\hat{\phi}$, $\hat{\psi}$,
$\hat{F}$ and $\lambda$, whose explicit forms are governed by Eq
(\ref{eq:promotion}).
Starting from a chiral field $\hat{\varphi}$ in the linear theory, one has
\begin{equation}
\int d^{4}xd^{2}\theta\partial_{\mu}\hat{\varphi}=0.
\end{equation}
Rewriting this identity in terms of nonlinear fields, we have
\begin{equation}
\int d^{4}xd^{2}\theta\det{T}\det{\emph{M}_{+}}\:\triangle_{\mu}^{+}\varphi=0.\label{eq:nl-chiral}
\end{equation}
Being a matter field in the nonlinear realization, each component
field of $\varphi$ transforms into itself and is independent of other
fields. Thus Eq (\ref{eq:nl-chiral}) contains actually three independent identities
\begin{equation}
\left\{ \begin{array}{l}
\int d^{4}xd^{2}\theta\det{T}\det{\emph{M}_{+}}\:\triangle_{\mu}^{+}(\phi)=0,\\
\int d^{4}xd^{2}\theta\det{T}\det{\emph{M}_{+}}\:\triangle_{\mu}^{+}(\sqrt{2}\theta\psi)=0,\\
\int d^{4}xd^{2}\theta\det{T}\det{\emph{M}_{+}}\:\triangle_{\mu}^{+}(F\theta^{2})=0.
\end{array}\right.
\end{equation}
 Explicitly, one has
\begin{equation}
\int d^{4}x\det{T}\:4\kappa^{2}\left(\bar{\lambda}_{\rho}\bar{\sigma}^{\nu\rho}\bar{\lambda}_{\nu}\nabla_{\mu}\phi+2\bar{\lambda}_{\mu}\bar{\sigma}^{\nu\rho}\bar{\lambda}_{\rho}\nabla_{\nu}\phi\right)=0,\label{eq:chiral 1}
\end{equation}
\begin{equation}
\int d^{4}x\det{T}\:\sqrt{2}\kappa\left[i(\nabla_{\mu}\psi\sigma^{\nu}\bar{\lambda}_{\nu}-\nabla_{\nu}\psi\sigma^{\nu}\bar{\lambda}_{\mu})+4\kappa^{2}(\bar{\lambda}_{\rho}\bar{\sigma}^{\nu\rho}\bar{\lambda}_{\nu}\lambda_{\mu}\psi+2\bar{\lambda}_{\mu}\bar{\sigma}^{\nu\rho}\bar{\lambda}_{\rho}\lambda_{\nu}\psi)\right]=0,\label{eq:chiral 2}
\end{equation}
\begin{equation}
\int d^{4}x\det{T}\left[\nabla_{\mu}F+2i\kappa^{2}F\left(\lambda_{\mu}\sigma^{\nu}\bar{\lambda}_{\nu}-\lambda_{\nu}\sigma^{\nu}\bar{\lambda}_{\mu}\right)\right]=0.\label{eq:chiral 3}
\end{equation}
We now have three complicated integrands which are total-divergences. 
The corresponding integrals vanish and do not contribute to any physical process.
Having obtained these identities, the explicit forms of $\phi$, $\psi$ and $F$ are irrelevant.
$\phi$, $\psi$ and $F$ could be any fundamental or composite fields.
From the linear identity $\int d^{4}xd^{2}\bar{\theta}\partial_{\mu}\hat{\varphi}^{\dagger}=0$,
three identities about $\phi^{\dagger}$, $\bar{\psi}$ and $F^{\dagger}$ can be obtained. They are just conjugate identities
of Eq (\ref{eq:chiral 1}), Eq(\ref{eq:chiral 2}) and Eq (\ref{eq:chiral 3}),
respectively.

More identities can be obtained following the same spirit.\footnote{Identities with spinor indices in nonlinear theories can be derived
from non-linearizing $\int d^{4}xd^{4}\theta D_{\alpha}\hat{\Phi}=0$ and $\int d^{4}xd^{4}\theta\bar{D}_{\dot{\alpha}}\hat{\Phi}=0$
in linear theories. Since they are not Lorentz scalars, thus of no help to constructing nonlinear actions.
We do not discuss them in this paper. }
There should be more identities if one considers a general super-field $\hat{\Phi}$ of nine independent component fields.
Nine new identities result from non-linearizing $\int d^{4}xd^{4}\theta\partial_{\mu}\hat{\Phi}=0$ and they will be listed in Appendix. 
As dual descriptions, those processes could be regarded as rewritten identities in the language of the nonlinear fields.
These identities are very complicated and hard to be obtained via other methods.
They all contain integrands that are total-divergences,
where matter fields can be either fundamental or composite.
Thus, no new identities can be obtained even if one starts with identities with more than one derivatives in the linear theory, such as $\int d^{4}xd^{4}\theta\partial_{\mu}\partial_{\nu}\hat{\Phi}=0$.

In summary, Eq (\ref{eq:identity G1}) only involves the Goldstino
field. A nonlinear complex scaler field $\phi$, not matter fundamental or composite, obeys Eq (\ref{eq:chiral 1}), Eq (\ref{eq:chiral 3}), Eq (\ref{eq:a7})
and Eq (\ref{eq:a10}). Correspondingly, a nonlinear complex scaler field $\phi^{\dagger}$
obeys the conjugate identities of them. A nonlinear complex fermion field $\psi$,
fundamental or composite, obeys Eq (\ref{eq:chiral 2}), Eq (\ref{eq:a4})
and Eq (\ref{eq:a9}). Conjugate identities of Eq (\ref{eq:chiral 2}),
Eq (\ref{eq:a4}) and Eq (\ref{eq:a9}) are concerning a nonlinear complex fermion
field $\bar{\psi}$. For a nonlinear vector boson field $\nu_{u}$, it should
obey the same identities as a scaler field should. Moreover, $\nu_{u}$
should obey Eq (\ref{eq:a5}) especially.

Our dual descriptions are only applicable to UV SUSY-breaking theories, 
which could be described either linearly or nonlinearly. 
When some superpartners have masses much higher than the typical energy scale of the concerned physical processes, they could be integrated out. 
In contrast to the linear realization, 
heavy particles can be integrated out without breaking SUSY in the nonlinear realization.
The low energy spectrum does not consist of complete supersymmetric multiplets, 
but all remaining fields respect the SUSY algebra in a nonlinear fashion. 
We cannot construct dual descriptions for low energy effective SUSY-breaking theories.
However, our identities are still valid for low energy effective SUSY actions.
This is because of the same algebraic structure of the nonlinear SUSY realization.
A general nonlinear effective action, which is SUSY and gauge invariant, is given as \cite{Clark:1996aw},\cite{Clark:1997aa},\cite{Klein:2002vu}
\begin{equation}
\mathcal{S}_{\mathrm{eff}}=\int d^{4}x\det{T}\:\mathcal{L}_{\mathrm{eff}}(\lambda_{\mu},\bar{\lambda}_{\mu},\phi^{i},\nabla_{\mu}\phi^{i},\mathcal{F}_{\mu\nu}).
\end{equation}
Here the effective Lagrangian $\mathcal{L}_{\mathrm{eff}}$ can be
expanded in the number of the Goldstino fields, namely $\mathcal{L}_{\mathrm{eff}}=\mathcal{L}_{(0)}+\mathcal{L}_{(1)}+\mathcal{L}_{(2)}+...$.
The leading-order term $\mathcal{L}_{(0)}$ is
fixed by the Lagrangian of MSSM while the high-order terms $\mathcal{L}_{(n)}$
are essentially model dependent. 
To be variant under nonlinear SUSY transformations, 
effective actions could be constructed order by order in the number of Goldstino fields. 
In principle our new identities can simplify $\mathcal{L}_{\mathrm{eff}}$ by finding the total-divergences,
which masquerade as non-zero results.
However, none such application has been seen because it is challenging to construct the most general $\mathcal{L}_{\mathrm{eff}}$ via this method.

As an application of these identities, 
we will prove presently the equivalence between the canonical Kahler potential ${\cal S}_{{\rm can}}^{{\rm NL}}$ and ${\cal S}_{{\rm can}}^{'{\rm NL}}$. 
Keeping the dual description in mind, ${\cal S}_{{\rm can}}^{{\rm NL}}$
and ${\cal S}_{{\rm can}}^{'{\rm NL}}$ should describe the same physics,
since their linear versions ${\cal S}_{{\rm can}}$ and ${\cal S}_{{\rm can}}^{'}$ do. 
As discussed before, ${\cal S}_{{\rm can}}^{{\rm NL}}=\int d^{4}xd^{4}\theta\det{\emph{T}}\det{\emph{M}}\;\Phi^{\dagger}\Phi$ while ${\cal S}_{{\rm can}}^{'{\rm NL}}=\int d^{4}xd^{4}\theta\frac{1}{2}\det{\emph{T}}\left(\det{\emph{M}_{+}}e^{-2i\theta\sigma^{\mu}\bar{\theta}\triangle_{\mu}^{-}}\varphi^{\dagger}\varphi+\det{\emph{M}_{-}}\varphi^{\dagger}e^{2i\theta\sigma^{\mu}\bar{\theta}\triangle_{\mu}^{+}}\varphi\right)$ 
in the nonlinear realization. Integrating out the $\theta$'s,
one has
\begin{eqnarray}
{\cal S}_{{\rm can}}^{{\rm NL}} & = & \int d^{4}x\det{\emph{T}}\:\left[F^{\dagger}F+\frac{1}{2}\phi\nabla^{\mu}\phi_{\mu}^{\dagger}+\frac{1}{2}\phi^{\dagger}\nabla^{\mu}\phi_{\mu}-\frac{1}{4}\nabla_{\mu}\left(\phi\phi^{\mu\dagger}\right)-\frac{1}{4}\nabla_{\mu}\left(\phi^{\dagger}\phi^{\mu}\right)\right.\label{eq:1}\\
 &  & +\frac{i}{2}\nabla_{\mu}\bar{\psi}\bar{\sigma}^{\mu}\psi-\frac{i}{2}\bar{\psi}\bar{\sigma}^{\mu}\nabla_{\mu}\psi+\frac{\kappa}{\sqrt{2}}\phi_{\mu}^{\dagger}\lambda_{\nu}\sigma^{\mu}\bar{\sigma}^{\nu}\psi+\frac{\kappa}{\sqrt{2}}\phi_{\mu}\bar{\lambda}_{\nu}\bar{\sigma}^{\mu}\sigma^{\nu}\bar{\psi}\nonumber \\
 &  & +\frac{\kappa^{2}}{2}\epsilon^{\mu\nu\rho\gamma}\lambda_{\mu}\sigma_{\gamma}\bar{\lambda}_{\nu}\left(\phi\phi_{\rho}^{\dagger}-\phi^{\dagger}\phi_{\rho}\right)-i\kappa^{2}\phi\phi_{\mu}^{\dagger}\lambda_{\nu}\sigma^{\mu}\bar{\sigma}^{\nu\rho}\bar{\lambda}_{\rho}+i\kappa^{2}\phi^{\dagger}\phi_{\mu}\lambda_{\rho}\sigma^{\rho\nu}\sigma^{\mu}\bar{\lambda}_{\nu}\nonumber \\
 &  & +\frac{\kappa}{\sqrt{2}}\bar{\lambda}_{\mu}\bar{\sigma}^{\mu\nu}\nabla_{\nu}\bar{\psi}\phi-\frac{\kappa}{\sqrt{2}}\bar{\lambda}_{\mu}\bar{\sigma}^{\mu\nu}\bar{\psi}\nabla_{\nu}\phi+\frac{\kappa}{\sqrt{2}}\lambda_{\mu}\sigma^{\mu\nu}\nabla_{\nu}\psi\phi^{\dagger}-\frac{\kappa}{\sqrt{2}}\lambda_{\mu}\sigma^{\mu\nu}\psi\nabla_{\nu}\phi^{\dagger}\nonumber \\
 &  & +\frac{i\kappa^{3}}{\sqrt{2}}\left(\bar{\lambda}_{\mu}\bar{\sigma}^{\mu\nu}\bar{\lambda}_{\nu}\lambda_{\rho}\sigma^{\rho}\bar{\psi}+2\bar{\lambda}_{\nu}\bar{\sigma}^{\rho\mu}\bar{\lambda}_{\rho}\lambda_{\mu}\sigma^{\nu}\bar{\psi}\right)\phi\nonumber \\
 &  & -\left.\frac{i\kappa^{3}}{\sqrt{2}}\left(\lambda_{\mu}\sigma^{\mu\nu}\lambda_{\nu}\psi\sigma^{\rho}\bar{\lambda}_{\rho}+2\lambda_{\nu}\sigma^{\rho\mu}\lambda_{\rho}\psi\sigma^{\nu}\bar{\lambda}_{\mu}\right)\phi^{\dagger}\right],\nonumber
\end{eqnarray}
\begin{eqnarray}
{\cal S}_{{\rm can}}^{'{\rm NL}} & = & \int d^{4}x\det{\emph{T}}\:\left[F^{\dagger}F+\frac{1}{2}\phi\nabla^{\mu}\phi_{\mu}^{\dagger}+\frac{1}{2}\phi^{\dagger}\nabla^{\mu}\phi_{\mu}\right.\label{eq:2}\\
 &  & +\frac{i}{2}\nabla_{\mu}\bar{\psi}\bar{\sigma}^{\mu}\psi-\frac{i}{2}\bar{\psi}\bar{\sigma}^{\mu}\nabla_{\mu}\psi+\frac{\kappa}{\sqrt{2}}\phi_{\mu}^{\dagger}\lambda_{\nu}\sigma^{\mu}\bar{\sigma}^{\nu}\psi+\frac{\kappa}{\sqrt{2}}\phi_{\mu}\bar{\lambda}_{\nu}\bar{\sigma}^{\mu}\sigma^{\nu}\bar{\psi}\nonumber \\
 &  & -2i\kappa^{2}\phi\phi_{\mu}^{\dagger}\lambda_{\nu}\sigma^{\mu}\bar{\sigma}^{\nu\rho}\bar{\lambda}_{\rho}+2i\kappa^{2}\phi^{\dagger}\phi_{\mu}\lambda_{\rho}\sigma^{\rho\nu}\sigma^{\mu}\bar{\lambda}_{\nu}\nonumber \\
 &  & +\left.\sqrt{2}\kappa\bar{\lambda}_{\mu}\bar{\sigma}^{\mu\nu}\nabla_{\nu}\bar{\psi}\phi+\sqrt{2}\kappa\lambda_{\mu}\sigma^{\mu\nu}\nabla_{\nu}\psi\phi^{\dagger}\right],\nonumber
\end{eqnarray}
where $\phi_{\mu}=\nabla_{\mu}\phi+\sqrt{2}\kappa\psi\lambda_{\mu}$. 
Taking $\frac{1}{4}\phi\phi^{\mu\dagger}$ as a composite field and using Eq (\ref{eq:chiral 3}), 
one has
\begin{equation}
\int d^{4}x\det{T}\left[\frac{1}{4}\nabla_{\mu}\left(\phi\phi^{\mu\dagger}\right)\right]=\int d^{4}x\det{T}\left[-\frac{i\kappa^{2}}{2}\phi\phi^{\mu\dagger}\left(\lambda_{\mu}\sigma^{\nu}\bar{\lambda}_{\nu}-\lambda_{\nu}\sigma^{\nu}\bar{\lambda}_{\mu}\right)\right].\label{eq:prove1}
\end{equation}
Taking $\frac{i\kappa}{4\sqrt{2}}\phi(\sigma^{\mu}\bar{\psi})_{\alpha}$ as a composite fermion field, 
Eq (\ref{eq:chiral 2}) leads to
\begin{eqnarray}
 &  & \int d^{4}x\det{T}\left[\frac{i\kappa^{3}}{\sqrt{2}}\phi\left(\bar{\lambda}_{\rho}\bar{\sigma}^{\nu\rho}\bar{\lambda}_{\nu}\lambda_{\mu}\sigma^{\mu}\bar{\psi}+2\bar{\lambda}_{\mu}\bar{\sigma}^{\nu\rho}\bar{\lambda}_{\rho}\lambda_{\nu}\sigma^{\mu}\bar{\psi}\right)\right]\label{eq:prove2}\\
 & = & \int d^{4}x\det{T}\left[-\frac{\kappa}{4\sqrt{2}}\bar{\lambda}_{\nu}\bar{\sigma}^{\nu}\nabla_{\mu}\left(\sigma^{\mu}\bar{\psi}\phi\right)+\frac{\kappa}{4\sqrt{2}}\bar{\lambda}_{\mu}\bar{\sigma}^{\nu}\nabla_{\nu}\left(\sigma^{\mu}\bar{\psi}\phi\right)\right].\nonumber
\end{eqnarray}
Substituting Eq (\ref{eq:prove1}), Eq (\ref{eq:prove2}) and 
their conjugate identities back into ${\cal S}_{{\rm can}}^{{\rm NL}}$,
we can obtain ${\cal S}_{{\rm can}}^{'{\rm NL}}$, as expected. 
The difference between ${\cal S}_{{\rm can}}^{{\rm NL}}$
and ${\cal S}_{{\rm can}}^{'{\rm NL}}$ are indeed integrals over total divergences. 

When coupled to gauge fields, the canonical Kahler potential in linear
theory becomes ${\cal S}_{{\rm can}}^{\rm V}=\int d^{4}xd^{4}\theta\hat{\Phi}^{\dagger}e^{2g\hat{V}}\hat{\Phi}$,
where $\hat{V}$ is a vector superfield. The corresponding potential
in the standard nonlinear realization of SUSY is \cite{Ivanov:1982bpa}
\begin{equation}
{\cal S}_{{\rm can}}^{{\rm V,NL}}=\int d^{4}xd^{4}\theta\det{\emph{T}}\det{\emph{M}}\;\Phi^{\dagger}e^{2gV}\Phi.\label{eq:11}
\end{equation}
Here $V$ could be constructed from $\hat{V}$ by Eq (\ref{eq:promotion}).
$\hat{V}$ and $V$ cannot be put into the Wess-Zumino gauge
simultaneously. In this paper, $V$ is chosen to be in Wess-Zumino
gauge \cite{Luo:2010zp}. That is to say,
\begin{eqnarray}
V & = & -\theta\sigma^{\mu}\bar{\theta}v_{\mu}+i\theta^{2}\bar{\theta}\left(\bar{\chi}-\frac{1}{2}\kappa\bar{\sigma}^{\mu}\sigma^{\nu}\bar{\lambda}_{\mu}v_{\nu}\right)-i\bar{\theta}^{2}\theta\left(\chi-\frac{1}{2}\kappa\sigma^{\mu}\bar{\sigma}^{\nu}\lambda_{\mu}v_{\nu}\right)\nonumber \\
 &  & +\frac{1}{2}\theta^{2}\bar{\theta}^{2}\left(D-\kappa^{2}\lambda_{\mu}\sigma^{\nu}\bar{\lambda}^{\mu}v_{\nu}\right).
\end{eqnarray}
Ref. \cite{Luo:2010zp} tells us that this nonlinear action could be given in another form as 
\begin{equation}
{\cal S}_{{\rm can}}^{'\rm V,{\rm NL}}=\int d^{4}xd^{4}\theta\frac{1}{2}\det{\emph{T}}\left(\det{\emph{M}_{+}}e^{-2i\theta\sigma^{\mu}\bar{\theta}\triangle_{\mu}^{-}}\varphi^{\dagger}e^{2gU}\varphi+\det{\emph{M}_{-}}\varphi^{\dagger}e^{2gU}e^{2i\theta\sigma^{\mu}\bar{\theta}\triangle_{\mu}^{+}}\varphi\right),\label{eq:22}
\end{equation}
where $U=\exp(-i\theta\sigma^{\mu}\bar{\theta}\triangle_{\mu})V$.
${\cal S}_{{\rm can}}^{'\rm V,{\rm NL}}$ can be shown to be equivalent to ${\cal S}_{{\rm can}}^{\rm V,{\rm NL}}$
with the help of Eq (\ref{eq:prove1}), Eq (\ref{eq:prove2}) and
their conjugate identities, following the same step of identifying ${\cal S}_{{\rm can}}^{'{\rm NL}}$
with ${\cal S}_{{\rm can}}^{{\rm NL}}$.

Now we turn to a general Kahler potential. Without gauge couplings, a general Kahler
potential in linear SUSY theory can be written as ${\cal S}_{{\rm gen}}=\int d^{4}xd^{4}\theta\,\mathcal{K}(\hat{\Phi}^{\dagger},\hat{\Phi})$
or a different form ${\cal S}_{{\rm gen}}^{'}=\int d^{4}xd^{4}\theta\frac{1}{2}\left[\mathcal{K}(e^{-2i\theta\sigma^{\mu}\bar{\theta}\partial_{\mu}}\hat{\varphi}^{\dagger},\hat{\varphi})+\mathcal{K}(\hat{\varphi}^{\dagger},e^{2i\theta\sigma^{\mu}\bar{\theta}\partial_{\mu}}\hat{\varphi})\right]$ up to total divergence terms.
For ${\cal S}_{{\rm gen}}$, its nonlinear version is \cite{Luo:2010zp}
\begin{equation}
{\cal S}_{{\rm gen}}^{\rm NL}=\int d^{4}xd^{4}\theta\det{\emph{T}}\det{\emph{M}}\,\mathcal{K}(\Phi^{\dagger},\Phi),
\end{equation}
Taylor expanding $\mathcal{K}(\Phi^{\dagger},\Phi)$  in terms of $\theta$ and $\bar{\theta}$ and then using the substitution rules discussed in \cite{Luo:2010zp}, the nonlinear action can be obtained exactly as
\begin{eqnarray*}
{\cal S}_{{\rm gen}}^{\rm NL} & = & \int d^{4}xd^{4}\theta\det{\emph{T}}\left\{ \frac{1}{4}\nabla^{\mu}\phi_{i\mu}\frac{\partial\mathcal{K}(\phi^{\dagger},\phi)}{\partial\phi_{i}}+\frac{1}{4}\nabla^{\mu}\phi_{i\mu}^{\dagger}\frac{\partial\mathcal{K}(\phi^{\dagger},\phi)}{\partial\phi_{i}^{\dagger}}+\frac{1}{2}i\kappa^{2}\lambda^{\mu}\sigma^{\nu}\bar{\lambda}_{\mu}\phi_{i\nu}\frac{\partial\mathcal{K}(\phi^{\dagger},\phi)}{\partial\phi_{i}}\right.\\
 &  & -\frac{1}{2}i\kappa^{2}\lambda^{\mu}\sigma^{\nu}\bar{\lambda}_{\mu}\phi_{i\nu}^{\dagger}\frac{\partial\mathcal{K}(\phi^{\dagger},\phi)}{\partial\phi_{i}^{\dagger}}+\frac{\sqrt{2}}{2}\kappa\lambda_{\mu}\sigma^{\mu\nu}\nabla_{\nu}\psi_{i}\frac{\partial\mathcal{K}(\phi^{\dagger},\phi)}{\partial\phi_{i}}+\frac{\sqrt{2}}{2}\kappa\bar{\lambda}_{\mu}\bar{\sigma}^{\mu\nu}\nabla_{\nu}\bar{\psi}_{i}\frac{\partial\mathcal{K}(\phi^{\dagger},\phi)}{\partial\phi_{i}^{\dagger}}\\
 &  & +\frac{1}{2}i\kappa^{2}\lambda_{\nu}\sigma^{\nu}\bar{\sigma}^{\mu}\sigma^{\rho}\bar{\lambda}_{\mu}\phi_{i\rho}\frac{\partial\mathcal{K}(\phi^{\dagger},\phi)}{\partial\phi_{i}}+\frac{1}{2}i\kappa^{2}\bar{\lambda}_{\nu}\bar{\sigma}^{\nu}\sigma^{\mu}\bar{\sigma}^{\rho}\lambda_{\mu}\phi_{i\rho}^{\dagger}\frac{\partial\mathcal{K}(\phi^{\dagger},\phi)}{\partial\phi_{i}^{\dagger}}\\
 &  & -\frac{1}{2}\kappa^{2}\epsilon^{\mu\nu\rho\gamma}\lambda_{\mu}\sigma_{\gamma}\bar{\lambda}_{\nu}\phi_{i\rho}\frac{\partial\mathcal{K}(\phi^{\dagger},\phi)}{\partial\phi_{i}}+\frac{1}{2}\kappa^{2}\epsilon^{\mu\nu\rho\gamma}\lambda_{\mu}\sigma_{\gamma}\bar{\lambda}_{\nu}\phi_{i\rho}^{\dagger}\frac{\partial\mathcal{K}(\phi^{\dagger},\phi)}{\partial\phi_{i}^{\dagger}}\\
 &  & -\frac{\sqrt{2}}{2}i\kappa^{3}\frac{\partial\mathcal{K}(\phi^{\dagger},\phi)}{\partial\phi_{i}}\psi_{i}\left[\sigma^{\rho}\bar{\lambda}_{\rho}\lambda_{\mu}\sigma^{\mu\nu}\lambda_{\nu}+2\sigma^{\nu}\bar{\lambda}_{\mu}\lambda_{\nu}\sigma^{\rho\mu}\lambda_{\rho}\right]\\
 &  & -\frac{\sqrt{2}}{2}i\kappa^{3}\frac{\partial\mathcal{K}(\phi^{\dagger},\phi)}{\partial\phi_{i}^{\dagger}}\bar{\psi}_{i}\left[\bar{\sigma}^{\rho}\lambda_{\rho}\bar{\lambda}_{\mu}\bar{\sigma}^{\mu\nu}\bar{\lambda}_{\nu}+2\bar{\sigma}^{\nu}\lambda_{\mu}\bar{\lambda}_{\nu}\bar{\sigma}^{\rho\mu}\bar{\lambda}_{\rho}\right]\\
 &  & +\frac{1}{2}i\nabla_{\mu}\bar{\psi}_{i}\bar{\sigma}^{\mu}\psi_{j}\frac{\partial^{2}\mathcal{K}(\phi^{\dagger},\phi)}{\partial\phi_{i}^{\dagger}\partial\phi_{j}}+\frac{\sqrt{2}}{2}\kappa\lambda_{\mu}\sigma^{\nu}\bar{\sigma}^{\mu}\psi_{j}\phi_{i\nu}^{\dagger}\frac{\partial^{2}\mathcal{K}(\phi^{\dagger},\phi)}{\partial\phi_{j}\partial\phi_{i}^{\dagger}}+\frac{\sqrt{2}}{2}\kappa\bar{\lambda}_{\mu}\bar{\sigma}^{\nu}\sigma^{\mu}\bar{\psi}_{j}\phi_{i\nu}\frac{\partial^{2}\mathcal{K}(\phi^{\dagger},\phi)}{\partial\phi_{i}\partial\phi_{j}^{\dagger}}\\
 &  & -\frac{1}{2}i\bar{\psi}_{i}\bar{\sigma}^{\mu}\nabla_{\mu}\psi_{j}\frac{\partial^{2}\mathcal{K}(\phi^{\dagger},\phi)}{\partial\phi_{i}^{\dagger}\partial\phi_{j}}-\frac{\sqrt{2}}{4}\kappa\lambda_{\mu}\sigma^{\mu}\bar{\sigma}^{\nu}\psi_{j}\phi_{i\nu}^{\dagger}\frac{\partial^{2}\mathcal{K}(\phi^{\dagger},\phi)}{\partial\phi_{j}\partial\phi_{i}^{\dagger}}-\frac{\sqrt{2}}{4}\kappa\bar{\lambda}_{\mu}\bar{\sigma}^{\mu}\sigma^{\nu}\bar{\psi}_{j}\phi_{i\nu}\frac{\partial^{2}\mathcal{K}(\phi^{\dagger},\phi)}{\partial\phi_{i}\partial\phi_{j}^{\dagger}}\\
 &  & +\frac{1}{2}\kappa^{2}i\epsilon^{\mu\nu\rho\gamma}\lambda_{\mu}\sigma_{\gamma}\bar{\lambda}_{\nu}\bar{\psi}_{j}\bar{\sigma}_{\rho}\psi_{i}\frac{\partial^{2}\mathcal{K}(\phi^{\dagger},\phi)}{\partial\phi_{i}\partial\phi_{j}^{\dagger}}+F_{i}^{\dagger}F_{j}\frac{\partial^{2}\mathcal{K}(\phi^{\dagger},\phi)}{\partial\phi_{i}^{\dagger}\partial\phi_{j}}-\frac{1}{2}\phi_{i\mu}^{\dagger}\phi_{j}^{\mu}\frac{\partial^{2}\mathcal{K}(\phi^{\dagger},\phi)}{\partial\phi_{i}^{\dagger}\partial\phi_{j}}\\
 &  & +\frac{1}{4}\phi_{i\mu}\phi_{j}^{\mu}\frac{\partial^{2}\mathcal{K}(\phi^{\dagger},\phi)}{\partial\phi_{i}\partial\phi_{j}}+\frac{1}{2}\kappa^{2}\lambda_{\mu}\sigma^{\mu\nu}\lambda_{\nu}\psi_{i}\psi_{j}\frac{\partial^{2}\mathcal{K}(\phi^{\dagger},\phi)}{\partial\phi_{i}\partial\phi_{j}}+\frac{1}{2}\kappa^{2}\bar{\lambda}_{\mu}\bar{\sigma}^{\mu\nu}\bar{\lambda}_{\nu}\bar{\psi}_{i}\bar{\psi}_{j}\frac{\partial^{2}\mathcal{K}(\phi^{\dagger},\phi)}{\partial\phi_{i}^{\dagger}\partial\phi_{j}^{\dagger}}\\
 &  & +\frac{1}{4}\phi_{i\mu}^{\dagger}\phi_{j}^{\dagger\mu}\frac{\partial\mathcal{K}(\phi^{\dagger},\phi)}{\partial\phi_{i}^{\dagger}\partial\phi_{j}^{\dagger}}+\frac{\sqrt{2}}{4}\kappa\lambda_{\nu}\sigma^{\nu}\bar{\sigma}^{\mu}\psi_{j}\phi_{i\mu}\frac{\partial^{2}\mathcal{K}(\phi^{\dagger},\phi)}{\partial\phi_{i}\partial\phi_{j}}+\frac{\sqrt{2}}{4}\kappa\bar{\lambda}_{\nu}\bar{\sigma}^{\nu}\sigma^{\mu}\bar{\psi}_{j}\phi_{i\mu}^{\dagger}\frac{\partial^{2}\mathcal{K}(\phi^{\dagger},\phi)}{\partial\phi_{i}^{\dagger}\partial\phi_{j}^{\dagger}}\\
 &  & +\frac{1}{2}i\psi_{k}\sigma^{\mu}\bar{\psi}_{i}\phi_{l\mu}\frac{\partial^{3}\mathcal{K}(\phi^{\dagger},\phi)}{\partial\phi_{i}^{\dagger}\partial\phi_{k}\partial\phi_{l}}-\frac{1}{2}i\psi_{i}\sigma^{\mu}\bar{\psi}_{k}\phi_{l\mu}^{\dagger}\frac{\partial^{3}\mathcal{K}(\phi^{\dagger},\phi)}{\partial\phi_{i}\partial\phi_{k}^{\dagger}\partial\phi_{l}^{\dagger}}\\
 &  & +\frac{\sqrt{2}}{4}i\kappa\lambda_{\mu}\sigma^{\mu}\bar{\psi}_{i}\psi_{k}\psi_{l}\frac{\partial^{3}\mathcal{K}(\phi^{\dagger},\phi)}{\partial\phi_{i}^{\dagger}\partial\phi_{k}\partial\phi_{l}}-\frac{\sqrt{2}}{4}i\kappa\psi_{i}\sigma^{\mu}\bar{\lambda}_{\mu}\bar{\psi}_{k}\bar{\psi}_{l}\frac{\partial^{3}\mathcal{K}(\phi^{\dagger},\phi)}{\partial\phi_{i}\partial\phi_{k}^{\dagger}\partial\phi_{l}^{\dagger}}\\
 &  & -\left.\frac{1}{2}\bar{\psi}_{k}\bar{\psi}_{l}F_{i}\frac{\partial^{3}\mathcal{K}(\phi^{\dagger},\phi)}{\partial\phi_{i}\partial\phi_{k}^{\dagger}\partial\phi_{l}^{\dagger}}-\frac{1}{2}\psi_{i}\psi_{j}F_{k}^{\dagger}\frac{\partial^{3}\mathcal{K}(\phi^{\dagger},\phi)}{\partial\phi_{i}\partial\phi_{j}\partial\phi_{k}^{\dagger}}+\frac{1}{4}\psi_{i}\psi_{j}\bar{\psi}_{k}\bar{\psi}_{l}\frac{\partial^{4}\mathcal{K}(\phi^{\dagger},\phi)}{\partial\phi_{i}\partial\phi_{j}\partial\phi_{k}^{\dagger}\partial\phi_{l}^{\dagger}}\right\} .
\end{eqnarray*}
Here fields with indices in Latin letters $i$, $j$, $k$, $l$
stand for different components while fields with indices in Greek
letters are defined as $\phi_{\mu}=\nabla_{\mu}\phi+\sqrt{2}\kappa\psi\lambda_{\mu}$.

For ${\cal S}_{{\rm gen}}^{'}$, its nonlinear version is \cite{Luo:2010zp}
\begin{equation}
{\cal S}_{{\rm gen}}^{'{\rm NL}}=\int d^{4}xd^{4}\theta\frac{1}{2}\det{\emph{T}}\left[\det{\emph{M}_{+}}\mathcal{K}(e^{-2i\theta\sigma^{\mu}\bar{\theta}\triangle_{\mu}^{-}}\varphi^{\dagger},\varphi)+\det{\emph{M}_{-}}\mathcal{K}(\varphi^{\dagger},e^{2i\theta\sigma^{\mu}\bar{\theta}\triangle_{\mu}^{+}}\varphi)\right].
\end{equation}
After expanding, one has
\begin{eqnarray*}
{\cal S}_{{\rm gen}}^{'\rm NL} & = & \int d^{4}xd^{4}\theta\det{\emph{T}}\left\{ \frac{1}{2}\nabla^{\mu}\phi_{k\mu}\frac{\partial\mathcal{K}(\phi^{\dagger},\phi)}{\partial\phi_{k}}+\frac{1}{2}\nabla^{\mu}\phi_{k\mu}^{\dagger}\frac{\partial\mathcal{K}(\phi^{\dagger},\phi)}{\partial\phi_{k}^{\dagger}}\right.\\
 &  & +i\kappa^{2}\lambda^{\mu}\sigma^{\nu}\bar{\lambda}_{\mu}\phi_{k\nu}\frac{\partial\mathcal{K}(\phi^{\dagger},\phi)}{\partial\phi_{k}}-i\kappa^{2}\lambda^{\mu}\sigma^{\nu}\bar{\lambda}_{\mu}\phi_{k\nu}^{\dagger}\frac{\partial\mathcal{K}(\phi^{\dagger},\phi)}{\partial\phi_{k}^{\dagger}}\\
 &  & +\sqrt{2}\kappa\lambda_{\mu}\sigma^{\mu\nu}\nabla_{\nu}\psi_{k}\frac{\partial\mathcal{K}(\phi^{\dagger},\phi)}{\partial\phi_{k}}+\sqrt{2}\kappa\bar{\lambda}_{\mu}\bar{\sigma}^{\mu\nu}\nabla_{\nu}\bar{\psi}_{k}\frac{\partial\mathcal{K}(\phi^{\dagger},\phi)}{\partial\phi_{k}^{\dagger}}\\
 &  & +i\kappa^{2}\lambda_{\mu}\sigma^{\mu}\bar{\sigma}^{\nu}\sigma^{\rho}\bar{\lambda}_{\nu}\phi_{k\rho}\frac{\partial\mathcal{K}(\phi^{\dagger},\phi)}{\partial\phi_{k}}+i\kappa^{2}\bar{\lambda}_{\mu}\bar{\sigma}^{\mu}\sigma^{\nu}\bar{\sigma}^{\rho}\lambda_{\nu}\phi_{k\rho}^{\dagger}\frac{\partial\mathcal{K}(\phi^{\dagger},\phi)}{\partial\phi_{k}^{\dagger}}\\
 &  & +\frac{1}{2}i\nabla_{\mu}\psi_{k}\sigma^{\mu}\bar{\psi}_{i}\frac{\partial^{2}\mathcal{K}(\phi^{\dagger},\phi)}{\partial\phi_{k}\partial\phi_{i}^{\dagger}}-\frac{1}{2}i\psi_{i}\sigma^{\mu}\nabla_{\mu}\bar{\psi}_{k}\frac{\partial^{2}\mathcal{K}(\phi^{\dagger},\phi)}{\partial\phi_{i}\partial\phi_{k}^{\dagger}}+F_{i}F_{k}^{\dagger}\frac{\partial^{2}\mathcal{K}(\phi^{\dagger},\phi)}{\partial\phi_{i}\partial\phi_{k}^{\dagger}}\\
 &  & +\frac{\sqrt{2}}{2}\kappa\phi_{\nu}^{\dagger}\lambda_{\mu}\sigma^{\nu}\bar{\sigma}^{\mu}\psi_{i}\frac{\partial^{2}\mathcal{K}(\phi^{\dagger},\phi)}{\partial\phi_{i}\partial\phi_{k}^{\dagger}}+\frac{\sqrt{2}}{2}\kappa\phi_{\nu}\bar{\lambda}_{\mu}\bar{\sigma}^{\nu}\sigma^{\mu}\bar{\psi}_{i}\frac{\partial^{2}\mathcal{K}(\phi^{\dagger},\phi)}{\partial\phi_{k}\partial\phi_{i}^{\dagger}}\\
 &  & +\frac{1}{2}\phi_{k\mu}\phi_{l}^{\mu}\frac{\partial^{2}\mathcal{K}(\phi^{\dagger},\phi)}{\partial\phi_{k}\partial\phi_{l}}+\frac{1}{2}\phi_{k\mu}^{\dagger}\phi_{l}^{\dagger\mu}\frac{\partial^{2}\mathcal{K}(\phi^{\dagger},\phi)}{\partial\phi_{k}^{\dagger}\partial\phi_{l}^{\dagger}}\\
 &  & +\kappa^{2}\lambda_{\mu}\sigma^{\mu\nu}\lambda_{\nu}\psi_{k}\psi_{l}\frac{\partial^{2}\mathcal{K}(\phi^{\dagger},\phi)}{\partial\phi_{k}\partial\phi_{l}}+\kappa^{2}\bar{\lambda}_{\mu}\bar{\sigma}^{\mu\nu}\bar{\lambda}_{\nu}\bar{\psi}_{k}\bar{\psi}_{l}\frac{\partial^{2}\mathcal{K}(\phi^{\dagger},\phi)}{\partial\phi_{k}^{\dagger}\partial\phi_{l}^{\dagger}}\\
 &  & +\frac{\sqrt{2}}{2}\kappa\lambda_{\mu}\sigma^{\mu}\bar{\sigma}^{\nu}\psi_{k}\phi_{l\nu}\frac{\partial^{2}\mathcal{K}(\phi^{\dagger},\phi)}{\partial\phi_{k}\partial\phi_{l}}+\frac{\sqrt{2}}{2}\kappa\bar{\lambda}_{\mu}\bar{\sigma}^{\mu}\sigma^{\nu}\bar{\psi}_{k}\phi_{l\nu}^{\dagger}\frac{\partial^{2}\mathcal{K}(\phi^{\dagger},\phi)}{\partial\phi_{k}^{\dagger}\partial\phi_{l}^{\dagger}}\\
 &  & +\frac{\sqrt{2}}{4}i\kappa\lambda_{\mu}\sigma^{\mu}\bar{\psi}_{i}\psi_{k}\psi_{l}\frac{\partial^{3}\mathcal{K}(\phi^{\dagger},\phi)}{\partial\phi_{i}^{\dagger}\partial\phi_{k}\partial\phi_{l}}-\frac{\sqrt{2}}{4}i\kappa\psi_{i}\sigma^{\mu}\bar{\lambda}_{\mu}\bar{\psi}_{k}\bar{\psi}_{l}\frac{\partial^{3}\mathcal{K}(\phi^{\dagger},\phi)}{\partial\phi_{i}\partial\phi_{k}^{\dagger}\partial\phi_{l}^{\dagger}}\\
 &  & +\frac{1}{2}i\psi_{k}\sigma^{\mu}\bar{\psi}_{i}\phi_{l\mu}\frac{\partial^{3}\mathcal{K}(\phi^{\dagger},\phi)}{\partial\phi_{i}^{\dagger}\partial\phi_{k}\partial\phi_{l}}-\frac{1}{2}i\psi_{i}\sigma^{\mu}\bar{\psi}_{k}\phi_{l\mu}^{\dagger}\frac{\partial^{3}\mathcal{K}(\phi^{\dagger},\phi)}{\partial\phi_{i}\partial\phi_{k}^{\dagger}\partial\phi_{l}^{\dagger}}\\
 &  & -\left.\frac{1}{2}\bar{\psi}_{k}\bar{\psi}_{l}F_{i}\frac{\partial^{3}\mathcal{K}(\phi^{\dagger},\phi)}{\partial\phi_{i}\partial\phi_{k}^{\dagger}\partial\phi_{l}^{\dagger}}-\frac{1}{2}\psi_{i}\psi_{j}F_{k}^{\dagger}\frac{\partial^{3}\mathcal{K}(\phi^{\dagger},\phi)}{\partial\phi_{i}\partial\phi_{j}\partial\phi_{k}^{\dagger}}+\frac{1}{4}\psi_{i}\psi_{j}\bar{\psi}_{k}\bar{\psi}_{l}\frac{\partial^{4}\mathcal{K}(\phi^{\dagger},\phi)}{\partial\phi_{i}\partial\phi_{j}\partial\phi_{k}^{\dagger}\partial\phi_{l}^{\dagger}}\right\} .
\end{eqnarray*}
${\cal S}_{\rm gen}^{{\rm NL}}$ and ${\cal S}_{{\rm gen}}^{'{\rm NL}}$
should describe the same physical process in the nonlinear SUSY realization.
We utilize our new identities to find out the nonlinear total-divergence terms. 
Taking $\frac{1}{4}\frac{\partial\mathcal{K}(\phi^{\dagger},\phi)}{\partial\phi_{i}^{\dagger}}\phi^{\mu\dagger}$
as a composite field into Eq (\ref{eq:chiral 3}), one has
\begin{equation}
\int d^{4}x\det{T}\frac{1}{4}\nabla_{\mu}\left[\frac{\partial\mathcal{K}(\phi^{\dagger},\phi)}{\partial\phi_{i}^{\dagger}}\phi_{i}^{\mu\dagger}\right]=\int d^{4}x\det{T}\left[-\frac{i\kappa^{2}}{2}\frac{\partial\mathcal{K}(\phi^{\dagger},\phi)}{\partial\phi_{i}^{\dagger}}\phi_{i}^{\mu\dagger}\left(\lambda_{\mu}\sigma^{\nu}\bar{\lambda}_{\nu}-\lambda_{\nu}\sigma^{\nu}\bar{\lambda}_{\mu}\right)\right].\label{eq:prove3}
\end{equation}
 Taking $\frac{\sqrt{2}i\kappa}{8}(\sigma^{\mu}\bar{\psi}_{i})_{\alpha}\frac{\partial\mathcal{K}(\phi^{\dagger},\phi)}{\partial\phi_{i}^{\dagger}}$
as a composite fermion field, Eq (\ref{eq:chiral 2}) gives
\begin{eqnarray}
 &  & \int d^{4}x\det{T}\left[\frac{i\kappa^{3}}{\sqrt{2}}\frac{\partial\mathcal{K}(\phi^{\dagger},\phi)}{\partial\phi_{i}^{\dagger}}\left(\bar{\lambda}_{\rho}\bar{\sigma}^{\nu\rho}\bar{\lambda}_{\nu}\lambda_{\mu}\sigma^{\mu}\bar{\psi}+2\bar{\lambda}_{\mu}\bar{\sigma}^{\nu\rho}\bar{\lambda}_{\rho}\lambda_{\nu}\sigma^{\mu}\bar{\psi}\right)\right]\label{eq:prove4}\\
 & = & \int d^{4}x\det{T}\left\{ -\frac{\kappa}{4\sqrt{2}}\bar{\lambda}_{\nu}\bar{\sigma}^{\nu}\nabla_{\mu}\left[\sigma^{\mu}\bar{\psi}\frac{\partial\mathcal{K}(\phi^{\dagger},\phi)}{\partial\phi_{i}^{\dagger}}\right]+\frac{\kappa}{4\sqrt{2}}\bar{\lambda}_{\mu}\bar{\sigma}^{\nu}\nabla_{\nu}\left[\sigma^{\mu}\bar{\psi}\frac{\partial\mathcal{K}(\phi^{\dagger},\phi)}{\partial\phi_{i}^{\dagger}}\right]\right\} .\nonumber
\end{eqnarray}
Substituting Eq (\ref{eq:prove3}), Eq (\ref{eq:prove4}) and their
conjugate identities back into ${\cal S}_{\mathrm{gen}}^{{\rm NL}}$,
we obtain ${\cal S}_{{\rm gen}}^{'{\rm NL}}$ from ${\cal S}_{\mathrm{gen}}^{{\rm NL}}$.
For a general Kahler potential coupled to a gauge field, the situation
is similar. The proof is still based on Eq (\ref{eq:prove3}), Eq
(\ref{eq:prove4}) and their conjugate identities.
Moreover, Eq (\ref{eq:prove3}) and Eq (\ref{eq:prove4}) can be exactly identical to Eq (\ref{eq:prove1}) and Eq (\ref{eq:prove2}) if let $\mathcal{K}(\hat{\Phi}^{\dagger},\hat{\Phi})=\hat{\Phi}^{\dagger}\hat{\Phi}$. 

Besides the standard nonlinear realization, 
there are already other formalisms that realize SUSY algebra nonlinearly. 
Keeping the dual description in mind, 
the total-divergence terms can be easily obtained in these formalisms. 
We take the chiral nonlinear realization as an example. 
In chiral nonlinear realization, the explicit SUSY transformations of the Goldstino
field $\tilde{\lambda}$ and matter fields $\tilde{\zeta}$ are \cite{Ivanov:1978mx}, \cite{Zumino:1974qc},  \cite{Samuel:1982uh}
\begin{equation}
\begin{cases}
\delta_{\xi}\tilde{\lambda}_{\alpha} & =\frac{\xi_{\alpha}}{\kappa}-2i\kappa\tilde{\lambda}\sigma^{\mu}\bar{\xi}\partial_{\mu}\tilde{\lambda}_{\alpha},\\
\delta_{\xi}\tilde{\zeta} & =-2i\kappa\tilde{\lambda}\sigma^{\mu}\bar{\xi}\partial_{\mu}\tilde{\zeta}.
\end{cases}
\end{equation}
In principle, both linear realization and this chiral nonlinear realization
can form dual descriptions of the same UV SUSY-breaking theory. 
The total-derivatives in this chiral nonlinear formalism could be obtained from
ones in the linear realization. 
But such a construction is not needed in practice, 
since it is well-known that the chiral fields $\tilde{\lambda}$
and $\tilde{\zeta}$ can be constructed from $\lambda$ and $\zeta$
in terms of 
\begin{equation}
\tilde{\lambda}_{\alpha}(x)=\lambda_{\alpha}(\omega),\qquad\tilde{\zeta}(x)=\zeta(\omega),
\end{equation}
with $\omega=x-i\kappa^{2}\lambda(\omega)\sigma\bar{\lambda}(\omega)$.
So all the identities about $\lambda$, $\zeta$ can be converted
into identities about $\tilde{\lambda}$, $\tilde{\zeta}$ by changing
of variables and adding the Jacobian determinant of the transformation. 
This means that different nonlinear realizations could form dual descriptions of the same theory.
For other nonlinear realizations, 
explicit maps between them and the standard nonlinear realization has been exactly given in \cite{Kuzenko:2011tj}. 
After changes of variables, 
the total-derivatives in these nonlinear formalisms could be obtained from
ones in the standard nonlinear realization. 
Finally, we pay attention to the constrained superfield formalism,
which has been proposed to construct nonlinear fields and actions \cite{Rocek:1978nb}, \cite{Lindstrom:1979kq}, \cite{Komargodski:2009rz}. 
The constrained superfield formalism could be regarded as a regular linear realization with some constraints.
The total-derivatives in the constrained formalism are the same as ordinary linear ones. 
The constraints only require that some component fields are not fundamental, but
consist of other component fields. 

\begin{acknowledgments}
This work is supported in part by the National Science Foundation
of China (10875103, 11135006) and National Basic Research Program
of China (2010CB833000).
\end{acknowledgments}
\appendix

\section{More Identities in Standard Nonlinear SUSY Realization}

For a general super-field $\hat{\Phi}=\hat{\phi}+\theta\hat{f}+\bar{\theta}\hat{\bar{g}}+\hat{m}\theta^{2}+\hat{n}\bar{\theta}^{2}+\theta\sigma^{\delta}\bar{\theta}\hat{\nu}_{\delta}+\bar{\theta}^{2}\theta\hat{\chi}+\theta^{2}\bar{\theta}\hat{\bar{\psi}}+\hat{D}\theta^{2}\bar{\theta}^{2}$ in linear theory,
we always have  $\int d^{4}xd^{4}\theta\partial_{\mu}\hat{\Phi}=0$. In terms of nonlinear fields, one has
\begin{equation}
\int d^{4}xd^{4}\theta\det{T}\det{\emph{M}}\triangle_{\mu}\Phi=0,\label{eq:a1}
\end{equation}
where $\triangle_{\mu}={(\emph{M}^{-1})_{\mu}}^{\nu}(\nabla_{\nu}+\kappa\lambda_{\nu}^{\alpha}\partial_{\alpha}+\kappa\bar{\lambda}_{\nu\dot{\alpha}}\partial^{\dot{\alpha}})$
and $\Phi=\phi+\theta f+\bar{\theta}\bar{g}+m\theta^{2}+n\bar{\theta}^{2}+\theta\sigma^{\delta}\bar{\theta}\nu_{\delta}+\bar{\theta}^{2}\theta\chi+\theta^{2}\bar{\theta}\bar{\psi}+D\theta^{2}\bar{\theta}^{2}$.
The matter fields in $\Phi$ are composites of component fields in
$\hat{\Phi}$ and the Goldstino field $\lambda$. Their explicit forms
are governed by Eq (\ref{eq:promotion}).

Defining
\[
\det{\emph{M}}{(\emph{M}^{-1})}_{\mu}^{\nu}=\delta_{\mu}^{\nu}+\theta\psi_{\mathcal{\emph{M}}\mu}^{\nu}+\bar{\theta}\bar{\psi}_{\mathcal{\emph{M}}\mu}^{\nu}+\theta\sigma_{\gamma}\bar{\theta}\nu_{\mathcal{\emph{M}}\mu}^{\gamma\nu}+\theta^{2}F_{\mathcal{\emph{M}}\mu}^{\nu}+\bar{\theta}^{2}F_{\mathcal{\emph{M}}\mu}^{\nu\dagger}+\theta^{2}\bar{\theta}\bar{\chi}_{\mathcal{\emph{M}\mu}}^{\nu}+\bar{\theta}^{2}\theta\chi_{\mathcal{\emph{M}}\mu}^{\nu}+\theta^{2}\bar{\theta}^{2}D_{\emph{M}\mu}^{\nu},
\]
 where
\[
\psi_{\mathcal{\emph{M}}\mu\alpha}^{\nu}=i\kappa\left[(\sigma^{\nu}\bar{\lambda}_{\mu})_{\alpha}-(\sigma^{\rho}\bar{\lambda}_{\rho})_{\alpha}\delta_{\mu}^{\nu}\right],
\]
\[
\nu_{\mathcal{\emph{M}}\mu}^{\gamma\nu}=i\kappa^{2}\left[\epsilon^{\nu\gamma\rho\delta}(\lambda_{\rho}\sigma_{\delta}\bar{\lambda}_{\mu}-\lambda_{\mu}\sigma_{\delta}\bar{\lambda}_{\rho})+\epsilon^{\gamma\sigma\rho\delta}\lambda_{\rho}\sigma_{\delta}\bar{\lambda}_{\sigma}\delta_{\mu}^{\nu}\right],
\]
\[
F_{\mathcal{\emph{M}}\mu}^{\nu}=\kappa^{2}\left[2\bar{\lambda}_{\mu}\bar{\sigma}^{\nu\rho}\bar{\lambda}_{\rho}-\bar{\lambda}_{\rho}\bar{\sigma}^{\rho\sigma}\bar{\lambda}_{\sigma}\delta_{\mu}^{\nu}\right],
\]
\begin{eqnarray*}
\chi_{\mathcal{\emph{M}}\mu\alpha}^{\nu} & = & \frac{i}{2}\kappa^{3}\left[2(\sigma^{\rho}\bar{\lambda}_{\rho})_{\alpha}\lambda_{\sigma}\sigma^{\sigma\gamma}\lambda_{\gamma}\delta_{\mu}^{\nu}+4(\sigma^{\gamma}\bar{\lambda}_{\sigma})_{\alpha}\lambda_{\gamma}\sigma^{\rho\sigma}\lambda_{\rho}\delta_{\mu}^{\nu}\right.\\
 &  & -2(\sigma^{\nu}\bar{\lambda}_{\mu})_{\alpha}\lambda_{\rho}\sigma^{\rho\sigma}\lambda_{\sigma}+4(\sigma^{\rho}\bar{\lambda}_{\mu})_{\alpha}\lambda_{\rho}\sigma^{\nu\sigma}\lambda_{\sigma}+4(\sigma^{\nu}\bar{\lambda}_{\rho})_{\alpha}\lambda_{\mu}\sigma^{\rho\sigma}\lambda_{\sigma}\\
 &  & +\left.(\sigma^{\sigma}\bar{\lambda}_{\sigma})_{\alpha}\lambda_{\mu}\sigma^{\rho}\bar{\sigma}^{\nu}\lambda_{\rho}-(\sigma^{\sigma}\bar{\lambda}_{\rho})_{\alpha}\lambda_{\mu}\sigma^{\rho}\bar{\sigma}^{\nu}\lambda_{\sigma}+i\epsilon^{\gamma\sigma\rho\delta}(\sigma_{\delta}\bar{\sigma}^{\nu}\lambda_{\mu})_{\alpha}\lambda_{\rho}\sigma_{\gamma}\bar{\lambda}_{\sigma}\right],
\end{eqnarray*}
\begin{eqnarray*}
D_{\emph{M}\mu}^{\nu} & = & \frac{1}{4}\kappa^{4}\left[4\lambda_{\sigma}\sigma^{\gamma\nu}\lambda_{\gamma}\bar{\lambda}_{\rho}\bar{\sigma}^{\sigma}\sigma^{\rho}\bar{\lambda}_{\mu}+4\lambda_{\mu}\sigma^{\rho}\bar{\sigma}^{\sigma}\lambda_{\rho}\bar{\lambda}_{\gamma}\bar{\sigma}^{\nu\gamma}\bar{\lambda}_{\sigma}\right.\\
 &  & -4\lambda_{\mu}\sigma^{\nu}\bar{\sigma}^{\gamma}\lambda_{\sigma}\bar{\lambda}_{\gamma}\bar{\sigma}^{\rho\sigma}\bar{\lambda}_{\rho}-4\bar{\lambda}_{\sigma}\bar{\sigma}^{\gamma}\sigma^{\nu}\bar{\lambda}_{\mu}\lambda_{\gamma}\sigma^{\rho\sigma}\lambda_{\rho}\\
 &  & +8\lambda_{\sigma}\sigma^{\sigma\gamma}\lambda_{\gamma}\bar{\lambda}_{\mu}\bar{\sigma}^{\rho\nu}\bar{\lambda}_{\rho}+8\bar{\lambda}_{\sigma}\bar{\sigma}^{\sigma\gamma}\bar{\lambda}_{\gamma}\lambda_{\mu}\sigma^{\rho\nu}\lambda_{\rho}\\
 &  & +\lambda_{\rho}\sigma^{\sigma}\bar{\sigma}^{\nu}(\lambda_{\gamma}\bar{\lambda}_{\sigma}-\lambda_{\sigma}\bar{\lambda}_{\gamma})\bar{\sigma}^{\gamma}\sigma^{\rho}\bar{\lambda}_{\mu}+\lambda_{\mu}\sigma^{\rho}\bar{\sigma}^{\gamma}(\lambda_{\sigma}\bar{\lambda}_{\gamma}-\lambda_{\gamma}\bar{\lambda}_{\sigma})\bar{\sigma}^{\nu}\sigma^{\sigma}\bar{\lambda}_{\rho}\\
 &  & +\lambda_{\mu}\sigma^{\rho}\bar{\sigma}^{\nu}(\lambda_{\gamma}\bar{\lambda}_{\sigma}-\lambda_{\sigma}\bar{\lambda}_{\gamma})\bar{\sigma}^{\gamma}\sigma^{\sigma}\bar{\lambda}_{\rho}+\lambda_{\rho}\sigma^{\sigma}\bar{\sigma}^{\gamma}(\lambda_{\sigma}\bar{\lambda}_{\gamma}-\lambda_{\gamma}\bar{\lambda}_{\sigma})\bar{\sigma}^{\nu}\sigma^{\rho}\bar{\lambda}_{\mu}\\
 &  & +\left.i(\lambda_{\mu}\sigma^{\rho}\bar{\sigma}_{\delta}\sigma^{\nu}\bar{\lambda}_{\rho}+\lambda_{\rho}\sigma^{\nu}\bar{\sigma}_{\delta}\sigma^{\rho}\bar{\lambda}_{\mu})\epsilon^{\sigma\gamma\delta\epsilon}\lambda_{\sigma}\sigma_{\epsilon}\bar{\lambda}_{\gamma}\right].
\end{eqnarray*}
Being a matter field in the nonlinear realization, each component
field of $\Phi$ transforms into itself and is independent of other
fields. Thus Eq (\ref{eq:a1})  contains nine independent identities. They are:
\begin{equation}
\int d^{4}x\det{T}\left[\nabla_{\mu}D+2i\kappa^{2}D(\lambda_{\mu}\sigma^{\nu}\bar{\lambda}_{\nu}-\lambda_{\nu}\sigma^{\nu}\bar{\lambda}_{\mu})\right]=0,\label{eq:a2}
\end{equation}
\begin{equation}
\int d^{4}x\det{T}(\kappa F_{\mathcal{\emph{M}}\mu}^{\nu+}\bar{\lambda}_{\nu}\bar{\psi}-\frac{1}{2}\bar{\psi}_{\mathcal{\emph{M}}\mu}^{\nu}\nabla_{\nu}\bar{\psi}+\frac{1}{2}\kappa v_{\mathcal{\emph{M}}\mu}^{\gamma\nu}\lambda_{\nu}\sigma_{\gamma}\bar{\psi})=0,\label{eq:a3}
\end{equation}
\begin{equation}
\int d^{4}x\det{T}(\kappa F_{\mathcal{\emph{M}}\mu}^{\nu}\lambda_{\nu}\chi-\frac{1}{2}\psi_{\mathcal{\emph{M}}\mu}^{\nu}\nabla_{\nu}\chi+\frac{1}{2}\kappa v_{\mathcal{\emph{M}}\mu}^{\gamma\nu}\chi\sigma_{\gamma}\bar{\lambda}_{\nu})=0,\label{eq:a4}
\end{equation}
\begin{equation}
\int d^{4}x\det{T}\left[v_{\mathcal{\emph{M}}\mu}^{\delta\nu}\nabla_{\nu}\nu_{\delta}+\kappa\nu_{\delta}(\lambda_{\nu}\sigma^{\delta}\bar{\chi}_{\mathcal{\emph{M}\mu}}^{\nu}+\chi_{\mathcal{\emph{M}}\mu}^{\nu}\sigma^{\delta}\bar{\lambda}_{\nu})\right]=0,\label{eq:a5}
\end{equation}
\begin{equation}
\int d^{4}x\det{T}(F_{\mathcal{\emph{M}}\mu}^{\nu}\nabla_{\nu}n-\kappa n\bar{\chi}_{\mathcal{\emph{M}\mu}}^{\nu}\bar{\lambda}_{\nu})=0,\label{eq:a6}
\end{equation}
\begin{equation}
\int d^{4}x\det{T}(F_{\mathcal{\emph{M}}\mu}^{\nu\dagger}\nabla_{\nu}m-\kappa m\chi_{\mathcal{\emph{M}}\mu}^{\nu}\lambda_{\nu})=0,\label{eq:a7}
\end{equation}
\begin{equation}
\int d^{4}x\det{T}(\kappa D_{\emph{M}\mu}^{\nu}\bar{\lambda}_{\nu}\bar{g}-\frac{1}{2}\bar{\chi}_{\mathcal{\emph{M}\mu}}^{\nu}\nabla_{\nu}\bar{g})=0,\label{eq:a8}
\end{equation}
\begin{equation}
\int d^{4}x\det{T}(\kappa D_{\emph{M}\mu}^{\nu}\lambda_{\nu}f-\frac{1}{2}\chi_{\mathcal{\emph{M}}\mu}^{\nu}\nabla_{\nu}f)=0,\label{eq:a9}
\end{equation}
\begin{equation}
\int d^{4}x\det{T}D_{\emph{M}\mu}^{\nu}\nabla_{\nu}\phi=0.\label{eq:a10}
\end{equation}
Noticing that Eq (\ref{eq:a2}) is exactly identical with Eq (\ref{eq:chiral 3}).
Eq (\ref{eq:a3}), Eq (\ref{eq:a6}) and Eq (\ref{eq:a8}) are conjugate
identities of Eq (\ref{eq:a4}), Eq (\ref{eq:a7}) and Eq (\ref{eq:a9})
respectively.

\bibliographystyle{aipnum4-1}
\bibliography{identities}

\end{document}